\documentclass[sigconf]{acmart}

\AtBeginDocument{%
  }

\copyrightyear{2025}
\acmYear{2025}
\setcopyright{cc}
\setcctype{by-nc}
\acmConference[UbiComp Companion '25]{Companion of the the 2025 ACM International Joint Conference on Pervasive and Ubiquitous Computing}{October 12--16, 2025}{Espoo, Finland}
\acmBooktitle{Companion of the the 2025 ACM International Joint Conference on
Pervasive and Ubiquitous Computing (UbiComp Companion '25), October 12--16, 2025, Espoo, Finland}\acmDOI{10.1145/3714394.3750586} \acmISBN{979-8-4007-1477-1/2025/10}

\begin{document}

\title{\emph{NieNie}: Adaptive Rhythmic System for Stress Relief with LLM-Based Guidance}


\author{Yichen Yu}
\authornote{Both authors contributed equally to this research.}
\orcid{0009-0001-0175-3253}

\affiliation{%
  \institution{Department of Computer Science}
  \institution{North Carolina State University}
  \city{Raleigh}
  \state{North Carolina}
  \country{USA}
}
\affiliation{%
  \institution{Georgia Institute of Technology}
  \city{Atlanta}
  \state{Georgia}
  \country{USA}
}
\email{lunarsboy@gmail.com}
\author{Qiaoran Wang}
\authornotemark[1]
\orcid{0009-0006-4483-1107}
\affiliation{%
  \institution{College of Design and Engineering}
  \institution{National University of Singapore}
  \city{Singapore}
  \country{Singapore}
}
\email{violetforever999@gmail.com}

\renewcommand{\shortauthors}{Yichen Yu \& Qiaoran Wan}

\begin{abstract}
Today's young people are facing increasing psychological stress due to various social issues. Traditional stress management tools often rely on static scripts or passive content, which are ineffective in alleviating stress. NieNie addresses this gap by combining rhythm biofeedback with real-time psychological guidance through a large language model (LLM), offering an interactive, tactile response. The system is specifically designed for young people experiencing emotional stress, collecting physiological signals such as heart rate variability and generating adaptive squeeze-release rhythms via soft, tactile devices. Utilising LLM, the system provides timely squeezing rhythms and psychologically guided feedback prompts, offering personalised rhythm games while reinforcing stress restructuring. Unlike traditional mental health apps, NieNie places users within an embodied interactive loop, leveraging tactile interaction, biofeedback, and adaptive language support to create an immersive stress regulation experience. This study demonstrates how embodied systems can connect bodily actions with mental health in everyday contexts.
\end{abstract}




\begin{teaserfigure}
  \includegraphics[width=\textwidth]{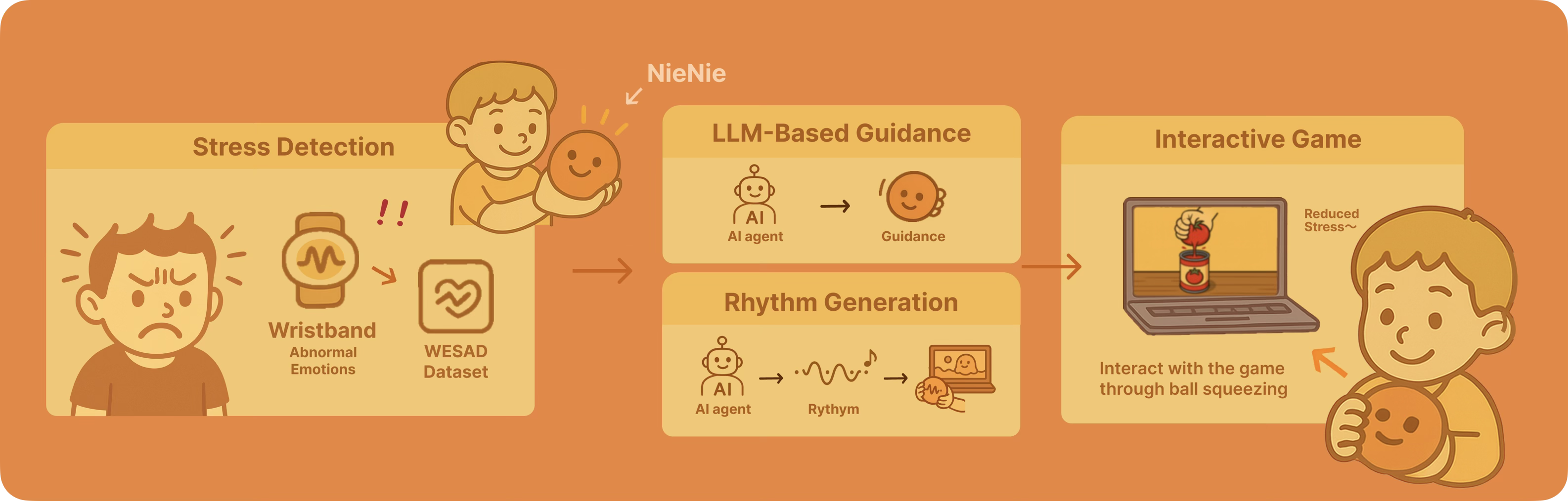}
  \caption{System overview of \emph{NieNie}, combining stress detection, LLM-driven rhythmic guidance, and interactive squeezing gameplay for emotional regulation.}
  \Description{Overview diagram showing NieNie system pipeline: Apple Watch stress detection, LLM-driven guidance, rhythmic squeeze-release feedback, and Unity game interaction.}
  \label{fig:teaser}
\end{teaserfigure}

\maketitle

\section{Introduction}
In recent years, mental health crises among college students have reached an alarming level. Students are facing increasing academic pressure, uncertain career prospects, social isolation, and growing digital distractions, all of which contribute to heightened psychological stress. Despite growing awareness and investment in mental health infrastructure, traditional intervention strategies such as mindfulness meditation apps, counseling sessions, and journaling prompts often fail to effectively engage students during moments of cognitive or emotional overload. Their core limitation lies in their reliance on user-driven engagement: individuals in a state of stress may lack the appropriate channels or methods to release stress. Additionally, most existing solutions remain static, lack personalization, and are disconnected from users' physical and emotional states during moments of stress \cite{burns2011mobile, gjoreski2018wearable}.

To address this significant gap, we have made \emph{NieNie}, a real-time, rhythm-guided system that senses physiological indicators of stress and guides users toward calmness through interactive gameplay. Our approach is grounded in research on affective computing, embodied interaction, and narrative psychology. Unlike treating stress as a cognitive challenge requiring rational solutions, \emph{NieNie} employs a more intuitive, body-based feedback loop. Users rhythmically squeeze a soft device under the system’s guidance, while the system responds with visual, auditory, and narrative cues that adaptively adjust based on the user’s physiological state. All guidance text and rhythms are not pre-written but generated in real-time by a large language model that simultaneously evaluates real-time sensor inputs and recent interaction history. This makes the experience smoother, more empathetic, and personalized, enabling the system to act as a dynamic co-regulator rather than a static guide \cite{gonzalez2020emotional, teixeira2023gaming}.

\emph{NieNie}’s core vision is to transform stress management from a cognitively demanding, text-heavy process into an embodied, emotionally resonant, and even playful experience. Unlike conventional breathing apps that provide countdown timers or voiceovers, \emph{NieNie} incorporates the user’s own sensorimotor feedback into the rhythm generation and response cycle. This not only helps reinforce positive behavioral loops but also fosters a sense of immersion and agency that is often missing from traditional wellness tools. Our system is designed with portability and scalability in mind, enabling deployment in all the places with minimal setup.

In the following sections, we will provide a detailed introduction to existing research related to this design, elaborate on the system architecture and implementation plan, and present the preliminary results of the pilot deployment. Finally, we will reflect on the current limitations of the system and propose future development directions and challenges.

\section{Related Work}
Our work on \emph{NieNie} builds upon a rich intersection of research in affective computing, biofeedback-based interaction, rhythmic entrainment, and personalized guidance via large language models. This convergence enables us to design a system that is not only reactive to stress but also engaging and adaptive to each user's unique emotional landscape.

Biofeedback games and stress sensing. Early explorations in biofeedback-based interaction have demonstrated the potential of physiological signals such as electrodermal activity (EDA), heart rate variability (HRV), and respiration to serve as real-time indicators of affective state \cite{gjoreski2018wearable, hernandez2014bioglass}. Notably, Schwartz et al.\cite{schwartz2017gbf} introduced a game-based biofeedback framework that incorporated EDA and HRV as continuous inputs for modulating gameplay difficulty. Their work showed that embedding physiological awareness into the mechanics of a game increased both user retention and emotional regulation efficacy. Subsequent research expanded this approach by exploring adaptive feedback loops that altered game dynamics based on biosignal fluctuations \cite{tammami2023adaptive}.

Rhythmic entrainment and embodied regulation. Teixeira et al.~\cite{teixeira2023gaming} demonstrated that rhythm-based breathing games on mobile devices significantly improved physiological indicators of calmness in users. Their work supports the broader theory of entrainment, wherein the human body and mind tend to synchronize with external rhythmic stimuli, such as pulsing lights, sounds, or haptic cues. In the context of \emph{NieNie}, we leverage this principle not through passive exposure but via active, rhythmic interaction—specifically, the act of squeezing and releasing a soft device in coordination with system generated timing cues. This embodied interaction aligns with research on somatic therapy and sensorimotor regulation, where bodily motion becomes a medium for emotional processing.

Narrative and emotional engagement. Emotional engagement is a key predictor of adherence and effectiveness in mental health interventions \cite{gonzalez2020emotional}. In \emph{NieNie}, we extend this by integrating real-time, adaptive narrative feedback generated by a large language model. Unlike scripted systems, our LLM-based module can adjust tone, style, and thematic elements based on the user’s ongoing emotional and physiological state, enabling a more responsive and personalized interaction loop.

LLMs for mental wellness. Valmeekam et al.\cite{valmeekam2023adaptive} demonstrate how LLMs can generate personalized prompts for emotional reflection and empathy. \emph{NieNie} positions this capability within an embodied gameplay context, bridging LLM-driven feedback with physical regulation strategies \cite{liang2023holistic, search3arxiv2023}.

\section{System Design}

\emph{NieNie} consists of four interlinked modules designed to detect stress and deliver personalized, rhythmic interventions through multimodal sensing and real-time feedback. Each module plays a distinct role while maintaining tight integration within a continuous feedback loop.

\textbf{Stress Detection.}  
At the foundation of \emph{NieNie} lies a real-time stress detection mechanism based on physiological sensing. Using an Apple Watch, the system captures heart rate and heart rate variability (HRV) via built-in PPG sensors. These signals are used to infer stress levels continuously during interaction, without requiring additional sensors or manual calibration. These three biomarkers commonly linked to autonomic nervous system activation. These raw signals are processed through an LSTM-based model trained on the WESAD dataset, which has been widely used for wearable stress classification \cite{gjoreski2018wearable}. The model operates on short, overlapping time windows and outputs a continuous stress probability score ranging from 0 (calm) to 1 (high stress). This probability is not treated as a fixed label but rather as a modulating signal that influences the downstream rhythm pacing and narrative tone. To accommodate inter-user variability, the system includes a lightweight calibration phase to adapt baseline thresholds to each user’s physiology. This ensures that feedback is not triggered unnecessarily, and that changes in physiological state rather than absolute values drive the adaptive logic of the experience.

\textbf{WESAD Dataset.}
To enable real-time stress inference in \emph{NieNie}, we trained a lightweight LSTM classifier using the WESAD dataset \cite{gjoreski2018wearable}, a benchmark for wearable-based stress detection. We focused on three physiological signals: electrodermal activity (EDA), skin temperature (TEMP), and heart rate (HR), sourced from the Empatica E4 sensor streams.

Each signal was preprocessed and truncated to equal length, then stacked into a multivariate time series. We segmented the data using a sliding window approach with a window size of 40 and a step size of 20, resulting in fixed-length input sequences of shape (40, 3). To simulate the three affective states—baseline, stress, and amusement—we assigned labels based on temporal segmentation of the recording sessions.

We trained a two-layer LSTM classifier implemented in PyTorch, using 128 hidden units and a final softmax output layer for three-class classification. The model was trained using cross-entropy loss and optimized with the Adam optimizer. A stratified 80/20 train-test split was used, and the model achieved strong performance on the held-out test set.

This LSTM classifier was later converted to TensorFlow Lite and quantized for efficient deployment on devices. During runtime, the model consumes real-time HRV and temperature data to produce a continuous probability estimate of user stress. This signal drives the dynamic adaptation of rhythm pacing and narrative tone within the \emph{NieNie} system.

\textbf{LLM-Based Guidance.}  
\emph{NieNie} uses large language models (LLMs) to generate real-time narrative prompts. Unlike systems that rely on fixed scripted messages, our LLM-driven module generates dynamic, personalized guidance based on rhythm adherence, pressure trajectories, and performance trends. For example, when a user maintains a steady squeezing motion in sync with the rhythm, the system might respond: ``Keep up this steady rhythm, you're in sync with the rhythm. Imagine you're squeezing this fruit rhythmically.'' During periods of irregular input or increased pressure, it may offer supportive prompts like: ''Try slowing down to find your rhythm again." These messages are concise, context-aware, and seamlessly integrated into the game narrative. The tone of feedback (supportive, neutral, or motivational) is adjusted in real-time to ensure it is both useful and non-intrusive. By combining physiological signals with adaptive language generation, this component transforms raw sensor data into emotionally resonant and actionable feedback \cite{valmeekam2023adaptive, liang2023holistic}.

This feedback mechanism directly supports subsequent modules, allowing users to interact with soft, fruit-shaped virtual objects via squeezing gestures. Haptic metaphors reinforce rhythm regulation, creating a closed-loop system between emotional states, narrative cues, and embodied interaction.

\textbf{Rhythm Generation.}  
Based on the inferred stress level, \emph{NieNie} generates squeeze-release rhythm patterns to guide the user's tactile interaction. This module sets time rhythm instructions (e.g., ``squeeze 5 times continuously''), which are adjusted according to the user's current autonomic nervous system state. When stress levels rise, the rhythm may begin with shorter, faster cycles to match the user's physiological arousal state, then gradually lengthen the cycles to promote parasympathetic activation and relaxation. Unlike systems relying on static patterns, the rhythm generator is dynamically adjusted in real-time by the system's large language model (LLM) component, which analyzes trends in user engagement and physiological metrics. The generated rhythm serves as a non-verbal, embodied intervention consistent with research on synchronization and breath-based emotional regulation\cite{teixeira2023gaming, abbal2024meta}. Visual and auditory cues embedded within the Unity environment reinforce the rhythm, enabling users to synchronize their bodily actions with system guidance in a fluid and intuitive manner.

\begin{figure}[h]
    \centering
    \begin{minipage}{0.32\linewidth}
        \centering
        \includegraphics[width=\textwidth]{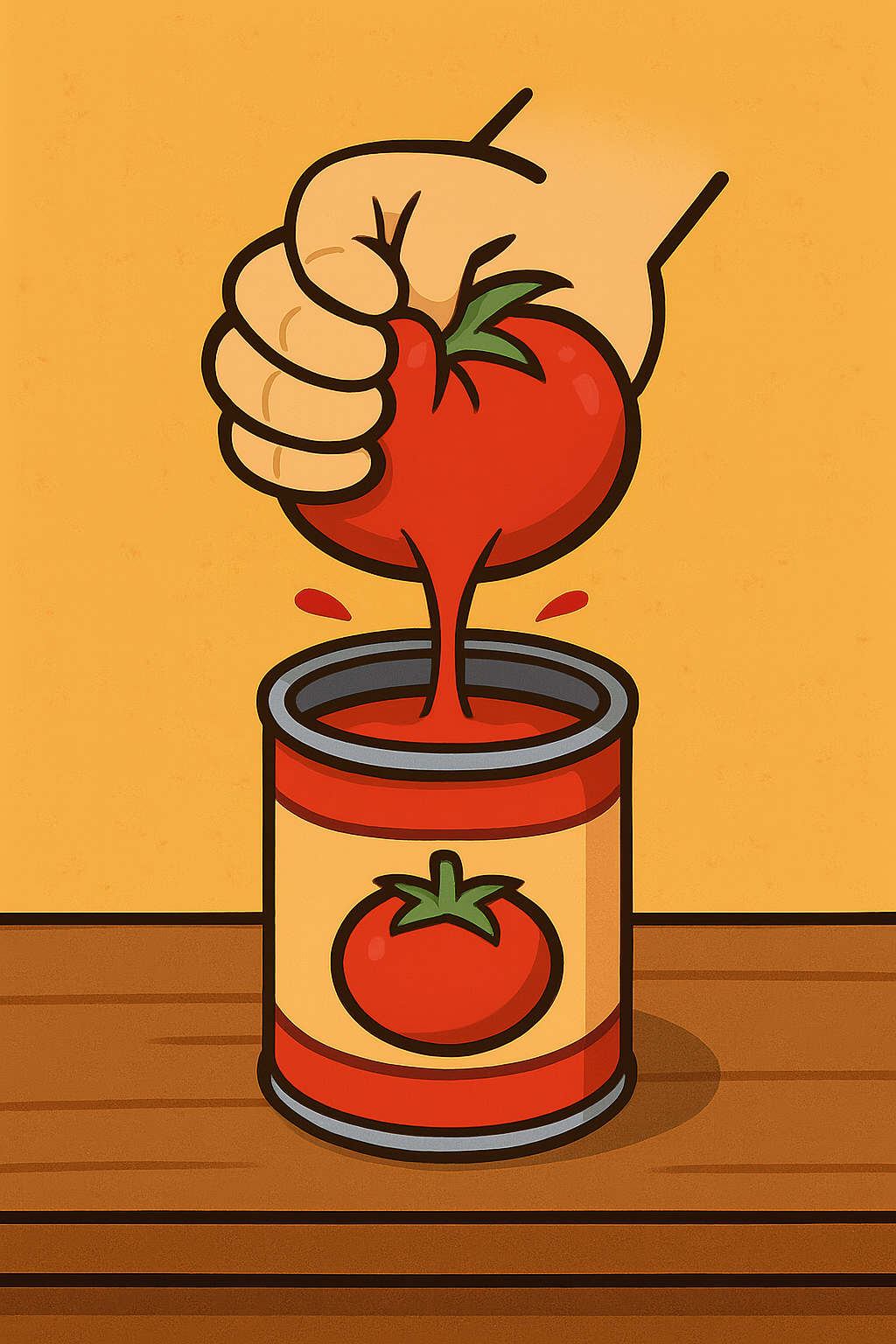}
    \end{minipage}
    \hfill
    \begin{minipage}{0.32\linewidth}
        \centering
        \includegraphics[width=\textwidth]{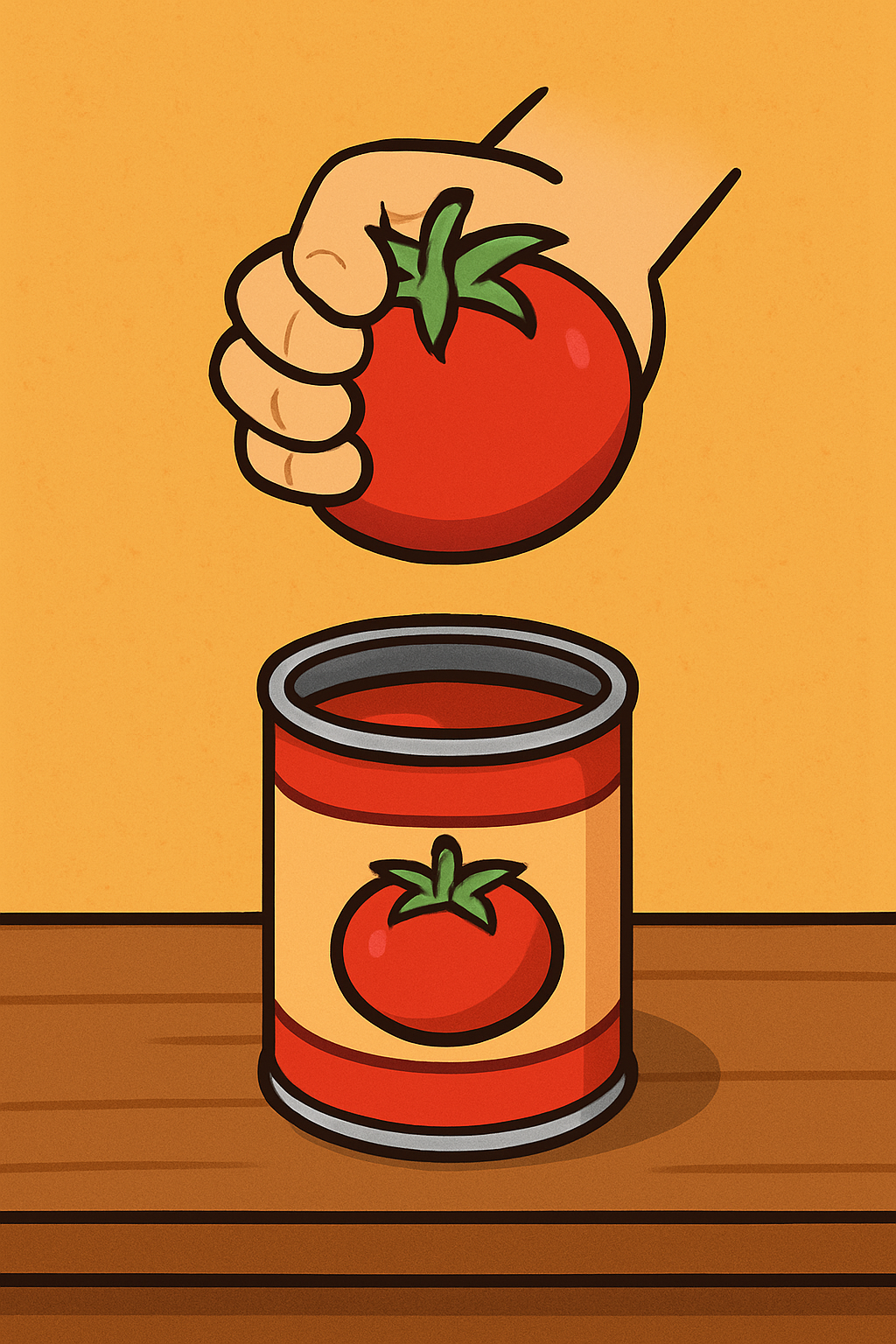}
    \end{minipage}
    \hfill
    \begin{minipage}{0.32\linewidth}
        \centering
        \includegraphics[width=\textwidth]{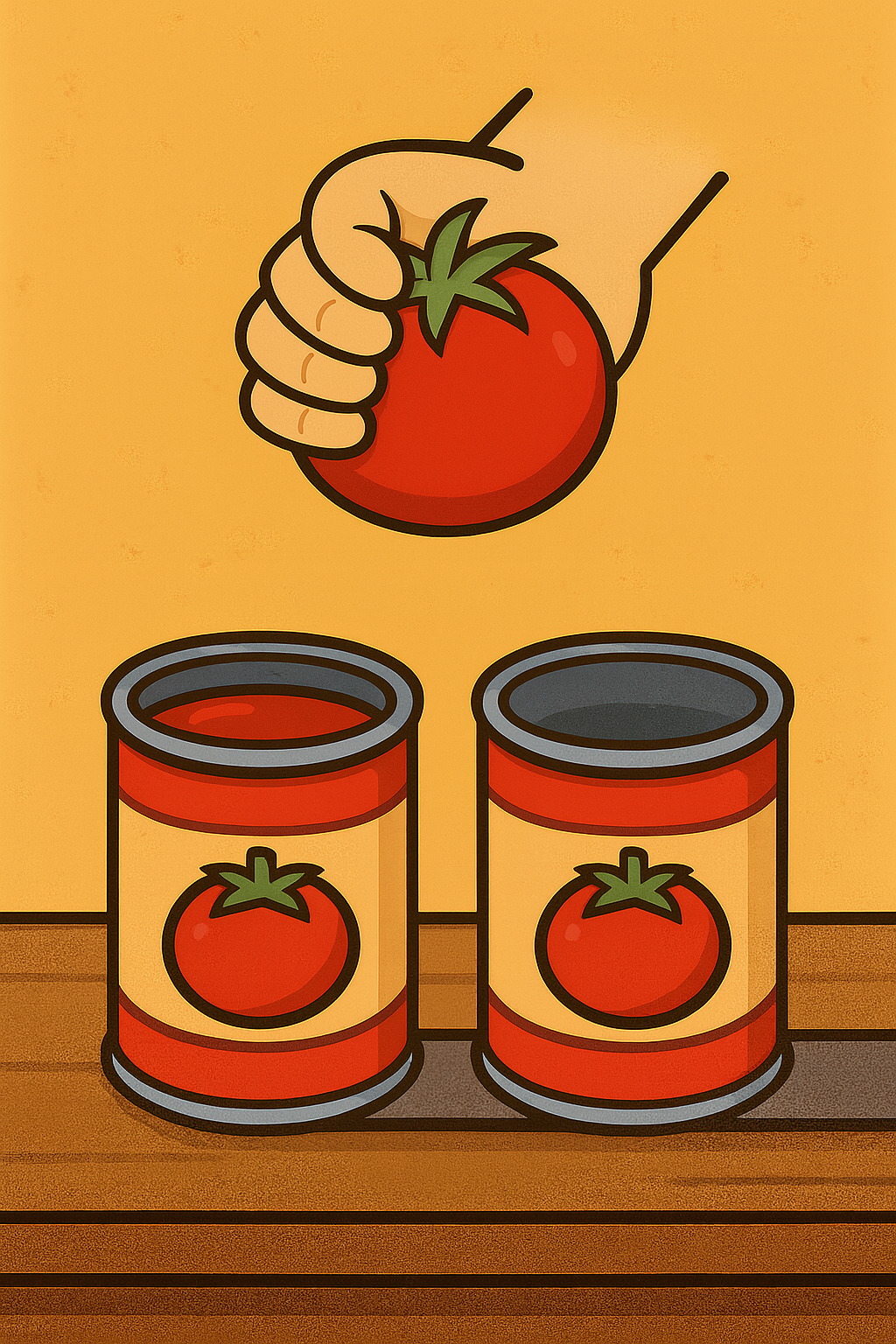}
    \end{minipage}
    \caption{Users squeeze tomatoes into the can according to the rhythm and frequency of the cans appearing. After squeezing, new tomatoes will appear. The filled cans slide to the left, while new empty cans slide in from the right.}
    \label{fig:tomato_examples}
 \end{figure}

\textbf{Interactive Game.}  
All feedback and perceptions converge in \emph{NieNie}'s Unity-based interactive game environment. Users interact by rhythmically squeezing a Bluetooth-connected soft device with a built-in pressure sensor. The game world responds in real time to the user's input and physiological state. Successful synchronization with the rhythm triggers positive in-game changes. These metaphors externalize users' internal states, helping them view their stress levels and no longer view them as hidden metrics but as malleable, interactive forces. For example, by following different in-game rhythm metaphors, users can squeeze in-game tomatoes to create tomato cans that meet rhythm requirements(see Figure~\ref{fig:tomato_examples}). Conversely, disconnection or imbalance encourages users to reengage. Unity handles all rendering and logic processing, synchronized with sensor inputs and language generated by large language models (LLMs). This integration creates a deeply immersive experience where every squeeze, pause, and narrative cue forms part of an ongoing feedback loop, supporting emotional regulation through presence, rhythm, and playfulness\cite{gonzalez2020emotional}.

\section{Implementation}
The \emph{NieNie} system is implemented using off-the-shelf wearable hardware, a custom-built soft interaction device, a Unity-based game engine, and a cloud-connected large language model API. This modular architecture enables real-time physiological sensing, multimodal feedback, and responsive gameplay within a lightweight mobile application \cite{hernandez2014bioglass, search6ml2024}.

\begin{figure}[h]
    \centering
    \includegraphics[width=0.45\textwidth]{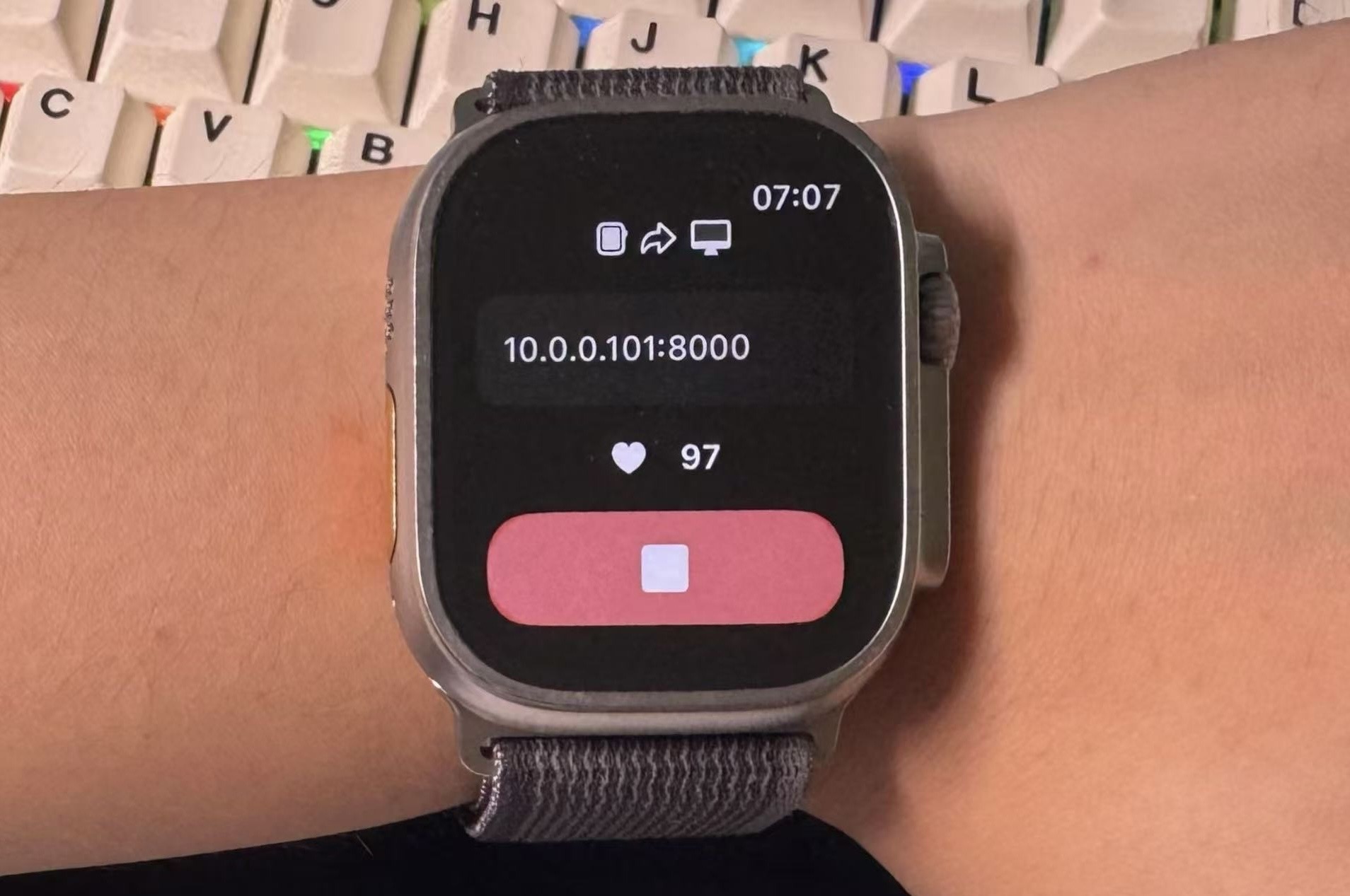}
    \caption{Apple Watch used in \emph{NieNie} for real-time heart rate and HRV sensing, with data transmitted to the Unity game system.}
    \label{fig:applewatch}
\end{figure}

Physiological data is collected using an Apple Watch, which directly provides heart rate and heart rate variability (HRV) measurements via built-in photoplethysmography (PPG) sensors(see Figure~\ref{fig:applewatch}). These metrics are streamed in real time to a paired iPhone using Apple's HealthKit and CoreBluetooth APIs, enabling continuous physiological monitoring without requiring manual signal processing or feature extraction. A compact LSTM classifier trained on the WESAD dataset continuously infers stress probabilities from overlapping time windows. The model is deployed via TensorFlow Lite for efficient on-device inference, achieving sub-30ms latency on modern smartphones \cite{ali2025teanet}.

The interaction system consists of a flexible pressure sensor embedded in a squeezable silicone sleeve. A microcontroller transmits normalized squeeze intensity values at 30Hz to Unity, which synchronizes the input with a rhythm clock that governs the pacing of squeeze–release cycles. This temporal alignment is used to calculate rhythm accuracy and engagement patterns in real time.

Unity renders an interactive game environment where users engage in rhythmic tasks such as squeezing virtual tomatoes in sync with a pacing rhythm. This visual interaction is complemented by natural language messages generated by an LLM, delivered through a backend server with optimized prompt templates. These messages are brief, adaptive, and emotionally supportive, enriching the experience without requiring textual input from the user.

The entire system runs as an application, ensuring portability and ease of deployment. Interaction logs, rhythm metrics, and LLM feedback history can be optionally recorded for further evaluation under IRB-compliant protocols.

\section{Limitations and Future Work}

Although \emph{NieNie} demonstrates the potential of integrating physiological sensing, rhythmic interaction, and feedback based on large language models (LLMs) in mobile systems, several limitations remain, which provide guidance for future research directions.

Accurately detecting stress from physiological data remains a challenge. Individual baseline differences, inconsistent sensor placement, and the transient nature of emotions can all impact inference accuracy. While current models utilize heart rate variability (HRV), their generalizability across broader populations has yet to be validated. Future research will explore user-specific calibration and the incorporation of additional sensor modalities to enhance detection.

The emotional consistency of generated feedback also introduces design complexity. While large language models can provide rich and adaptable information, they occasionally generate outputs that are contextually inconsistent, overly abstract, or emotionally inappropriate. Enhancing user autonomy (e.g., allowing message filtering or style customization) can help ensure more resonant and respectful interactions.
In terms of privacy and trust, even in systems that do not continuously store user data, continuous monitoring and personalized responses may raise concerns. Some users may perceive the system as intrusive or opaque, especially when content generated by LLMs lacks clear attribution.

Maintaining long-term engagement after the novelty wears off remains uncertain. While the system's gamified feedback loops can drive initial interaction, sustaining user motivation over the long term, especially in unstructured daily scenarios is a known challenge. Mechanisms such as goal setting, environmental variation, and narrative progression will be explored to promote long-term use and emotional impact.

Overall, addressing these limitations requires interdisciplinary collaboration across fields such as affective computing, human-computer interaction, behavioral psychology, and privacy research to ensure that \emph{NieNie} provides safe, inclusive, and meaningful solutions for mental health support in the real world.

\bibliographystyle{ACM-Reference-Format}
\bibliography{reference}

\end{document}